\documentclass[twocolumn, amssymb,pre]{revtex4}

\usepackage{graphicx}

\usepackage{float}

\newcommand{\etalp}{\textit{et~al.~}}

\begin{document}
\title{Inhomogeneous Relaxation Dynamics and Phase Behaviour of a {Liquid} Crystal Confined in a Nanoporous Solid}
\author{Sylwia~Ca{\l}us}
\affiliation{Faculty of Electrical Engineering, Czestochowa
University of Technology, 42-200 Czestochowa,
Poland}

\author{Andriy~V.~Kityk}
\affiliation{Faculty of Electrical Engineering, Czestochowa
University of Technology, 42-200 Czestochowa,
Poland}

\author{Manfred Eich}
\affiliation{Institute of Optical and Electronic Materials, Hamburg University of Technology (TUHH), D-21073 Hamburg-Harburg, Germany
}

\author{Patrick~Huber}
\affiliation{Institute of Materials Physics and Technology, Hamburg University of Technology (TUHH), D-21073 Hamburg-Harburg, Germany
}
\email[E-mail: ]{andriy.kityk@univie.ac.at, patrick.huber@tuhh.de}

\date{\today}

\begin{abstract}
We report filling-fraction dependent dielectric spectroscopy measurements on the relaxation dynamics of the rod-like nematogen 7CB condensed in 13~nm silica nanochannels. In the film-condensed regime, a slow interface relaxation dominates the dielectric spectra, whereas from the capillary-condensed state up to complete filling an additional, fast relaxation in the core of the channels is found. The temperature-dependence of the static capacitance, representative of the averaged, collective molecular orientational ordering, indicates a continuous, paranematic-to-nematic (P-N) transition, in contrast to the discontinuous bulk behaviour. It is well described by a Landau-de-Gennes free energy model for a phase transition in cylindrical confinement. The large tensile pressure of 10 MPa in the capillary-condensed state, resulting from the Young-Laplace pressure at highly curved liquid menisci, quantitatively accounts for a downward-shift of the P-N transition and an increased molecular mobility in comparison to the unstretched liquid state of the complete filling. The strengths of the slow and fast relaxations provide local information on the orientational order: The thermotropic behaviour in the core region is bulk-like, \textit{i.e.} it is characterized by an abrupt onset of the nematic order at the P-N transition. By contrast, the interface ordering exhibits a continuous evolution at the P-N transition. Thus, the phase behaviour of the entirely filled liquid crystal-silica nanocomposite can be quantitatively described by a linear superposition of these distinct nematic order contributions.
\end{abstract}

\maketitle

\section{Introduction}
\begin{figure}[h]
\begin{center}
\includegraphics[width=1 \linewidth, angle=0]{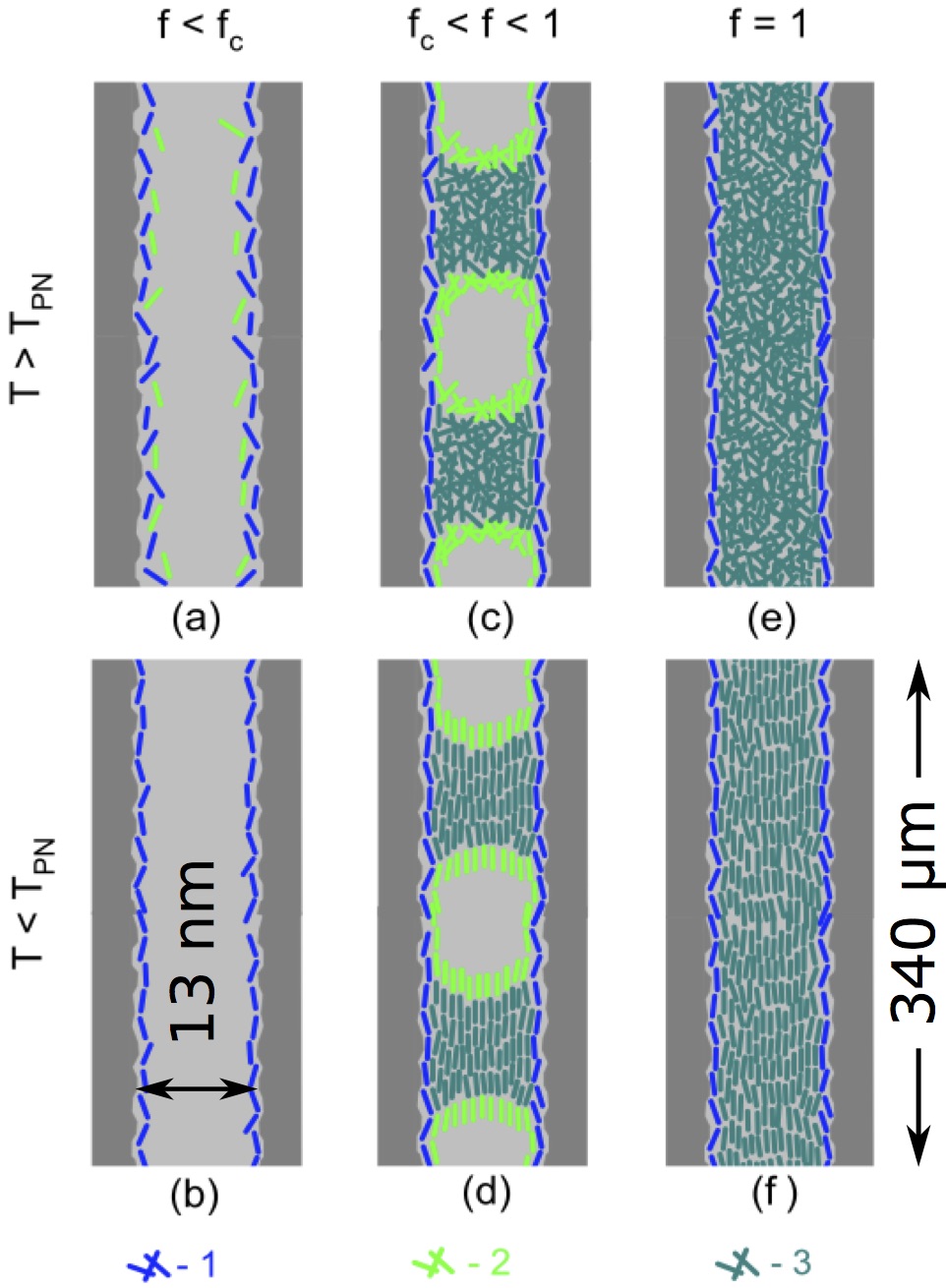} \caption{(color online). Schematic side view on distinct mesoscopic arrangements of a nematic liquid in a nano channel as a function of filling fraction $f$ above and below the paranematic-to-nematic transition point, $T_{PN}$: (a) and (b) refer to the film-condensed,  (c) and (d) to the capillary-condensed state and (e), (f) to the entirely filled channel. The critical filling fraction $f_c$ separates the film-condensed from the capillary-condensed state. The molecules in the interface layer (next to the pore walls), at the liquid-vapor interface and inside the core region of the channels are marked by different colors and labeled as species "1", "2" or "3".} \label{fig:structure}
\end{center}
\end{figure}

Liquid crystals (LCs) spatially confined on the nanometer scale exhibit structural and thermodynamical properties which differ often markedly from their bulk counterparts. Both the collective orientational (isotropic-to-nematic) and the translational (smectic-to-liquid or smectic-to-nematic) transitions have turned out to be significantly affected by finite size and interfacial (solid-liquid or liquid-liquid) interactions introduced by confining walls \cite{Ocko1990, Fukuto2008, Aya2011, Ruths2012, Beller2013, Lohr2014} or the geometrical constraints in nanoporous media \cite{Crawford1996, Huber2015, Bellini2001, Leheny2003, Shah2006, Kityk2008, Binder2008, Kityk2010, Grigoriadis2011, Araki2011, Kralj2012, Repnik2013, Lefort2014, Kityk2014a, Ndao2014, Zhang2014, Zhang2015, Karjalainen2015}. 

{For example, the second-order nematic-to-smectic-A transition is absent or greatly broadened in rod-like liquid crystals immersed in aerogels. This allowed a detailed study and comparison with theoretical predictions of the influence of quenched disorder introduced by random spatial confinement on the critical exponents of this prominent phase transition \cite{Crawford1996, Bellini2001, Repnik2013}. Similarly, spatial restrictions and interface interactions in mesoporous silica and alumina induce pronounced modifications of smectic ordering, e.g. often pre-smectic behaviour is observable far above the bulk nematic-to-smectic A transition. \cite{Crawford1996, Kralj2007}} It has also been shown experimentally \cite{Iannacchione1993, Kutnjak2003, Kityk2008}, in agreement with expectations from theory \cite{Crawford1996, Sheng1976}, that there is no ''true'' isotropic-nematic \textit{(I-N)} transition for LCs spatially restricted in at least one direction to a few nanometers. The molecular anchoring at the confining walls, quantified by an interfacial field, imposes a partial orientational, that results in a partially nematic ordering of the confined LCs, even at temperatures $T$ far above the bulk \textit{I-N} transition temperature $T^{\rm b}_{IN}$. The symmetry breaking and thus entropy change does not occur spontaneously, as characteristic of a genuine phase transition. It is enforced over relevant distances by the interaction with the walls. Thus confinement plays here a similar role as an external magnetic field for a spin system \cite{Stark2002, Repnik2013}: The strong first order \textit{I-N} transition is replaced by a weak first order or continuous paranematic-to-nematic (\textit{P-N}) transition at a temperature $T_{PN}$ and may also be accompanied by pre transitional phenomena in the molecular orientational distribution \cite{Eich1984}. An understanding of these phenomonenologies is of high fundament interest, for it allows to explore the validity (and break-down) of basic concepts of condensed matter science at the nanoscale \cite{Binder2008, Aya2011, Grigoriadis2011, Huber2015, Zhang2014}.

Moreover, LC confined in nanoporous solids provide promising composite materials for organic electronics and the emerging field of nano photonics, because of the mechanical rigidity of the solid host and the large variation in electrical and optical properties offered by the plethora of available LC systems along with the advent of tailorable nanoporous media \cite{Martin1994, Bisoyi2011, Abdulhalim2012, Duran2012}. 

Referring to the pioneering theoretical work of Sheng, Poniewierski, and Sluckin \cite{Sheng1976, Poniewierski1987} as well as  experimental work by Yokoyama\cite{Yokoyama1988} on LCs in semi-infinite, planar confinement, Kutnjak, Kralj, Lahajnar, and Zumer developed a Landau-de Gennes free energy ansatz \cite{Kutnjak2003,Kutnjak2004} (hereafter denoted as KKLZ model) for the \textit{I-N} transition in cylindrical pore geometry. This model describes the external orientational field as an effective \emph{nematic ordering field} $\sigma$. The KKLZ model very successfully describes the general features of the P-N~transition observed in spatial confinement \cite{Kutnjak2003, Kutnjak2004, Kityk2008, Calus2012, Huber2013,Calus2014}, in particular the $T$-dependence of the effective order parameter averaged over the entire pore geometry. However, it does not give any microscopic details on the orientational behaviour with regard to spatial inhomogeneities.

Computer simulations on LCs in thin film and pore geometry can give here important complementary insights \cite{Gruhn1997, Gruhn1998, Care2005, Binder2008, Ji2009, Ji2009b,Pizzirusso2012, Roscioni2013,Karjalainen2013, Cetinkaya2013,Schulz2014, Karjalainen2015}. These studies indicate pronounced spatial heterogeneities, in particular interface-induced molecular layering and radial gradients both in the orientational order and reorientational dynamics in cylindrical pore geometry \cite{Li2009}. Despite recent experimental advancements in optical techniques directly probing orientational order parameter profiles in the proximity of planar, solid walls \cite{Barna2008, DeLuca2008, Lee2009, Rosenblatt2014}, achieving the spatial (and temporal) resolution necessary to rigorously explore such inhomogeneities in nanometer-sized capillaries remains still experimentally extremely demanding, if not impossible. 

Dielectric spectroscopy on confined LCs, that is a measurement of the frequency-dependent complex electric capacity $C(\omega)$ can make important contributions in that respect \cite{Kremer2002}. This technique is based on the interaction of an external, alternating electric field with the electric dipole moments of the molecules. The molecules try to arrange their dipoles parallel to the extern field, however the resulting polarisation is disturbed by thermal noise and influenced by the interfacial orientational field. The time needed for dipoles to relax and to adjust to the alternating electrical field is sensitively depending on the local, collective mobility, most prominently the rotational viscosity \cite{Kremer2002}. Thus it sensitively affects the dielectric response, in particular the so-called ''$\delta$-relaxation process'' which  characterizes the movement of the permanent molecular dipole moment about the short geometrical axis of a dielectric positively anisotropic LC-molecule \cite{Birenheide1989}. 

A sizeable number of studies on LCs in porous media evidently document that the rate of dipolar relaxations (and thus orientational and translational mobility) usually differ significantly between the molecules in the pore wall proximity and the ones in the channel center \cite{Cramer1997, Hourri2001, Frunza2001, Leys2005, Sinha2005, Leys2008, Frunza2008, Bras2008, Jasiurkowska2012}. {In seminal dielectric studies (complemented by dynamic light scattering and calorimetric experiments) Aliev \etalp \cite{Sinha1997, Sinha1998,Aliev2005} could document a slow surface mobility in comparison to the dynamics in the pore centre for liquid crystals confined in tortuous and tubular mesopores. Moreover, a significant broadening of the dielectric spectra was found and traced by the authors to inhomogeneous couplings of the molecules to the pore walls and coupling variations among the molecules themselves \cite{Aliev2010}.} 

However, the detailed partitioning, the evolution of the relaxation behaviour and the superposition of the distinct relaxation contributions of LCs as a function of the filling of nanoporous media has been elusive so far and also controversially discussed \cite{Ji2009, Guegan2007, Lefort2008}. 
Here we employ dielectric relaxation spectroscopy on an archetypical rod-like liquid crystal (7 CB) confined in an array of isolated channels of approximately 7~nm radius in a monolithic silica membrane. Characteristic fillings (film condensate, capillary condensate and complete filling) - see illustrations in Fig. \ref{fig:structure}, which we documented in a previous optical birefringence and light scattering study on the identical system \cite{Huber2013}, are explored as a function of $T$. This along with the simple parallel capacitor geometry (see inset in Fig. \ref{fig:DSFillDependence}) allows us to derive detailed insights with regard to the heterogeneous mobility and orientational order of the confined molecular assemblies. Particularly, we can quantitatively describe the thermotropic behaviour with the KKLZ-model described above and show that the dielectric response of the organic-inorganic nano composite can be described by a \textit{linear} superposition of two distinct molecular populations, an interface species strongly affected in their molecular mobility and orientational order by the silica walls, and the collective molecular order in the pore centre.

\section{Experimental}
\subsection{Sample Preparation}
The nematogen liquid crystal 7CB (4-heptyl-4'-cyano biphenyl) has been purchased from Merck. The porous silicon membranes were prepared by electrochemical anodic etching of highly $p$-doped  $\langle$100$\rangle$ silicon wafers \cite{Canham1990, Lehmann1991, Sailor2011}. The samples were then subjected to thermal oxidation for 12 h at $T$=800 $^o$C under standard atmosphere. The resulting porous silica membranes are permeated by parallel channels aligned along the surface normal of the membrane \cite{Canham2015}. The average channel radius was $R=6.6 \pm$0.5 nm (porosity $P= 54 \pm$2\%) according to volumetric N$_2$-sorption isotherms recorded at $T$=77~K. 
\begin{figure}[h]
\begin{center}
  \includegraphics[width=1 \linewidth]{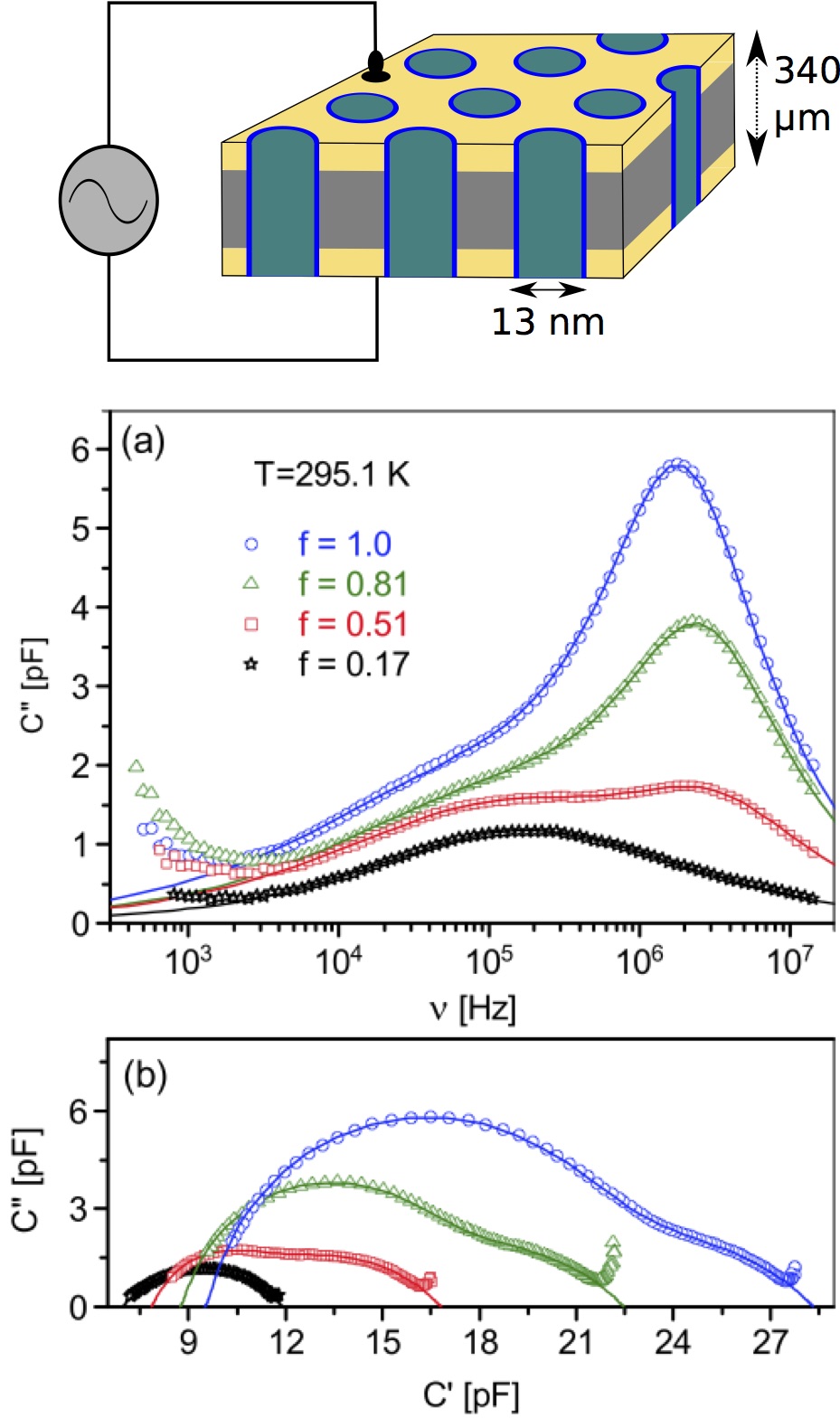}
  \caption{(top) Illustration of a dielectric relaxation experiment on a monolithic mesoporous membrane traversed by parallel nano channels filled with a liquid. The arrows indicate the nano channel's width and length, respectively. (a) Frequency dispersion of the imaginary part of the complex capacitance and (b) Cole-Cole plots for four different fractional fillings $f$ of 7CB in silica nanochannels ($R=6.6$ nm) at $T=295.1$ K. Symbols correspond to the experimental data points, whereas solid lines are fits with a two-relaxation model discussed in the text.} \label{fig:DSFillDependence}
\end{center}
\end{figure}

The entirely filled sample (fraction filling $f=$1.0) has been obtained by melt infiltration, that is capillarity-driven filling of the channels \cite{Huber2007, Gruener2009, Gruener2011, Huber2015}.  To prepare the partially filled samples ($0<f<1$) we imbibed the porous matrix with binary 7CB/cyclohexane solutions of selected concentrations. After evaporation of the high-vapor-pressure cyclohexane, the matrix remained filled by the low-vapor-pressure LC with a filling fraction $f$ that depends on the initial solute concentration. The filling fraction has been determined by weighting of the partially filled matrix and a comparison with the mass of the empty matrix. The evaporation of the solvent is controlled by time-dependent mass measurements of the sample at a temperature of 325~K and followed an exponential decay with time.  After 1 hour, the cyclohexane had evaporated to 1/100 of its initial concentration, and thus for the low-$f$ samples the further mass drop was below the detection limits of our balance, which was $5 \cdot 10^{-5}$ g. By a gradual increase of the solute concentration, we have prepared three filling fractions, $f=$ 0.17$\pm$0.01, 0.51$\pm$0.01 and 0.81$\pm$0.01 in addition to the completely filled sample, $f=$1.0. These selected fillings cover all known regimes of pore filling, as sketched in Fig.~\ref{fig:structure} and documented in a previous optical birefringence and light scattering study \cite{Huber2013}: the film-condensed regime ($f<f_c$), the capillary-condensed regime ($f_c <f<1$) and the complete filling ($f=1$). Note that according to our previous $f$-dependent optical data sets the critical filling fraction, separating the film-condensed and the capillary-condensed state, is $f_c \approx 0.25$ \cite{Huber2013}.

\subsection{Dielectric Relaxation Spectroscopy}
For the dielectric measurements, gold electrodes with thicknesses of 50 nm have been sputtered onto the porous membrane. All measurements were performed on samples that had an electrode area of 111 mm$^2$ and a thickness of 340 $\mu$m. The geometric capacitance of the samples is $C_0=2.89$~pF. Dielectric spectra have been recorded in the frequency range from 0.5~kHz to 15~MHz using the Impedance/Gain-Phase Analyzer Solatron-1260A. The measurements have been performed at 100 selected temperatures between 292~K and 324~K and the sample temperature was controlled with an accuracy of 0.01~K by a LakeShore 340 temperature controller.

\section{Results and Discussion}
\subsection{Molecular mobility upon channel filling probed by dielectric spectroscopy}

In Fig.~\ref{fig:DSFillDependence} we display the frequency dependencies of the imaginary part of the capacitance $C''(\nu)$ for four filling fractions at $T=$295.1~K. This is far below the P-N transition temperature $T_{PN}=315~K$, as determined in our previous birefringence study \cite{Huber2013}. $C''(\nu)$ is a measure of the frequency-dependent energy absorption of the nanocomposite, when the  $\delta$-relaxation occurs. Given the absence of any relaxation of the empty silica matrix in the frequency regime investigated, it represents a direct measure of the molecular relaxation behaviour of the confined LC molecules. Whereas the film-condensed regime ($f=$0.17) is characterized by a single, broad peak, indicating a  slow relaxation with a rather broad distribution of relaxation frequencies, another peak, typical of a fast relaxation process, evidently emerges on top of its high-frequency wing at higher filling fractions, $f$. It is characterized by a relatively narrow relaxation frequency distribution and its relaxation strength (height) rises with $f$ up to the entire filling. Thus, the series of dielectric dispersions shown in Fig.~\ref{fig:DSFillDependence} are quite instructive, since they document a partitioning of the pore filling in two components with distinct molecular mobility: (i) slowly relaxing molecules that are located in a direct contact with the pore walls (''interface species'', see Fig.~\ref{fig:structure}) and (ii) fast relaxing molecules that are located in the core region of the capillary bridges (a ''core'' component, see Fig.~\ref{fig:structure}). 

{In order to analyse the observed relaxation behaviour in more detail, we resort to a representation of the dielectric data in so-called Cole-Cole diagrams, i.e. we plot $C''(\nu)$ vs. $C'(\nu)$ with frequency $\nu$ as the independent parameter, see Fig. \ref{fig:DSFillDependence}c. In such a diagram, a material that has a single relaxation time $\tau$, as typical of the classical Debye relaxator, will appear as a semicircle with its centre lying on the horizontal (at $C''=0$) and the peak of the loss factor occurs at 1/$\tau$. Here, $\tau$ is a measure of the dipolar mobility, \textit{i.e.} it characterises the time required by the system to relax by thermally activated molecular motions to 1/$e$ times the dipolar ordering induced by the external electrical field. A material with a symmetric distribution of relaxation times will be a semicircle with its center lying below the horizontal at $C'(\nu)=0$. An asymmetric distribution of relaxation times, e.g. the Havriliak-Negami relaxation, results in an asymmetric arc. In that sense, the Cole-Cole representation allows one to geometrically illustrate and analyse the relaxation behaviour of a given system in a quite simple manner. \cite{Cole1941, Kremer2002}}

{As can be seen in Fig. \ref{fig:DSFillDependence}c the relaxation behaviour of our confined liquid crystalline system is characterized by one or two semicircles, depending on the filling fraction. Thus, its dielectric relaxation behaviour can be described by one or two, so-called Cole-Cole processes:
\begin{eqnarray}
C^*(\omega)&=&\varepsilon^*(\omega) C_0=\nonumber \\
&=&C_{\infty}+\frac{\Delta C_1}{1+(\textrm{i}\omega \tau_1)^{1-\alpha_1}}+\frac{\Delta C_2}{1+(\textrm{i}\omega \tau_2)^{1-\alpha_2}}, \quad
\label{eq:CFit}
\end{eqnarray}
Here $\omega=2\pi\nu$ is the cyclic frequency, $\varepsilon^*(\omega)$ is the complex dielectric constant of the composite, $C_{\infty}=\varepsilon_{\infty}C_0$ is the high frequency limit capacitance expressed via the high frequency permittivity $\varepsilon_{\infty}$, $\Delta C_1=\Delta \varepsilon_1 C_0$ and $\Delta C_{2}=\Delta \varepsilon_2 C_0$ are the capacitance relaxation strengths expressed via the dielectric relaxation strengths $\Delta\varepsilon_1$ and $\Delta \varepsilon_2$ of the slow process I and of the fast process II, respectively. The dielectric relaxation strengths quantify the amplitude of the different relaxation contributions, and thus characterise the extent of dipolar orientational polarization typical of each elementary process. The corresponding mean relaxation times are $\tau_1$ and $\tau_2$, respectively. 
The exponent parameter $\alpha$, can take values between 0 and 1, and allows one to describe different spectral shapes. When $\alpha=0$, the Cole-Cole model reduces to the Debye model (with a single relaxation frequency). When $\alpha>$0, the relaxation is stretched, i.e. it extends over a wider range on a logarithmic frequency scale than the Debye process.} 

{We will see below, that the spectral relaxation shapes (and thus $\alpha$ values) change in a characteristic manner as a function of temperature and filling fraction for our system. However, they are well represented by two symmetric Cole-Cole processes.} Deviations of the experimental data from the fitting curves at low frequencies can be attributed to ionic DC-conductivity. For this reason, the low-frequency region was always excluded from the fitting process. A special remark has to be made regarding the fitting procedure in the high temperature region (see dielectric spectra in the electronic supplementary), particularly just above $T_{PN}$, for the fractional fillings, $f=$0.51, 0.81 and 1.0. There, the maximum of the imaginary part, $C''(\nu)$ shifts out of the upper limit of the frequency window ($\nu_m>$15 MHz) employed in our experiment. Thus only a part of the left wing of the relaxation band is observed. This leads to an increased uncertainty of the extracted fit parameters, especially for the fast relaxation process II. Fortunately, one of the parameters affected, namely the capacitance relaxation strength  $\Delta C_2$, can be determined in an alternative way. Thereby uncertainties can be resolved in respect to all other fit parameters for this process. In the low-frequency limit ($\nu \rightarrow 0$) Eq.~\ref{eq:CFit} gives the static capacitance $C_{\mathrm{st}}$. It reads:
\begin{equation}
C_{\mathrm{st}}=C_{\infty}+\Delta C_1+\Delta C_2.
\label{eq:CStat}
\end{equation}
Since the nanochannels in the silica membrane are parallel-aligned to the applied electric field, the entirely filled sample forms a simple parallel circuit. Therefore, the effective permittivity of the composite is given as $\varepsilon = (1-P)\varepsilon_{\mathrm{SiO_2}}+P\varepsilon_{\mathrm{7CB}}$, where $\varepsilon_{\mathrm{SiO_2}} =$ 3.8 and $\varepsilon_{\mathrm{7CB}}$ are the permittivities of the silica host and 7CB guest, respectively. With a $C_{\infty}=$9.5 pF, as determined at 293 K from a Cole-Cole plot, one obtains $\varepsilon_{7CB}\approx 2.85$  by its extrapolation to high frequencies, \textit{i.e.} the value nearly equal to the dielectric permittivity of 7CB at optical frequencies. This means that the contribution of the LC component to the dielectric permittivity at high frequencies is due to electronic polarizability only, which is weakly temperature dependent.

\begin{figure}[H]
\begin{center}
\includegraphics[width=0.8 \linewidth]{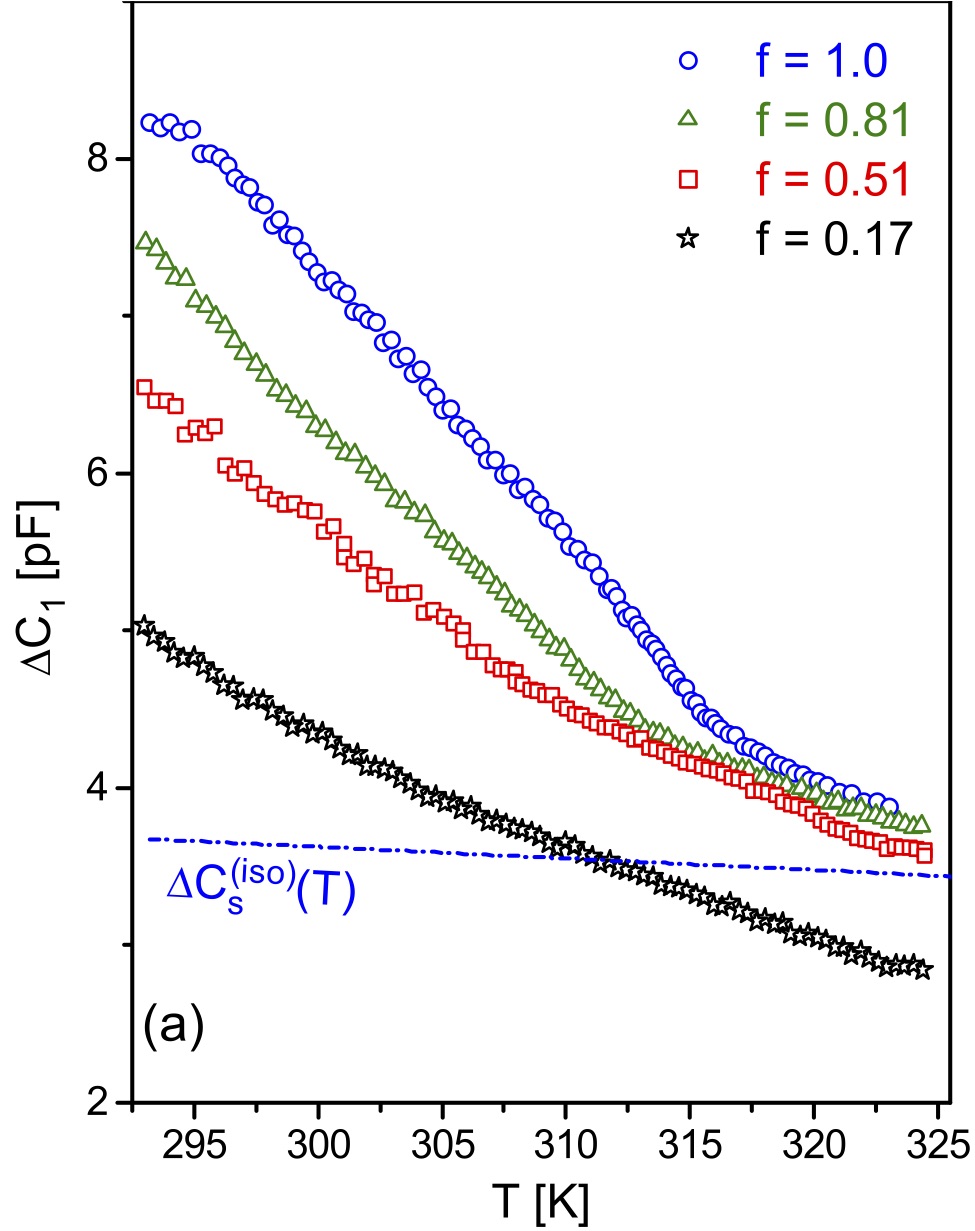}
\includegraphics[width=0.8 \linewidth]{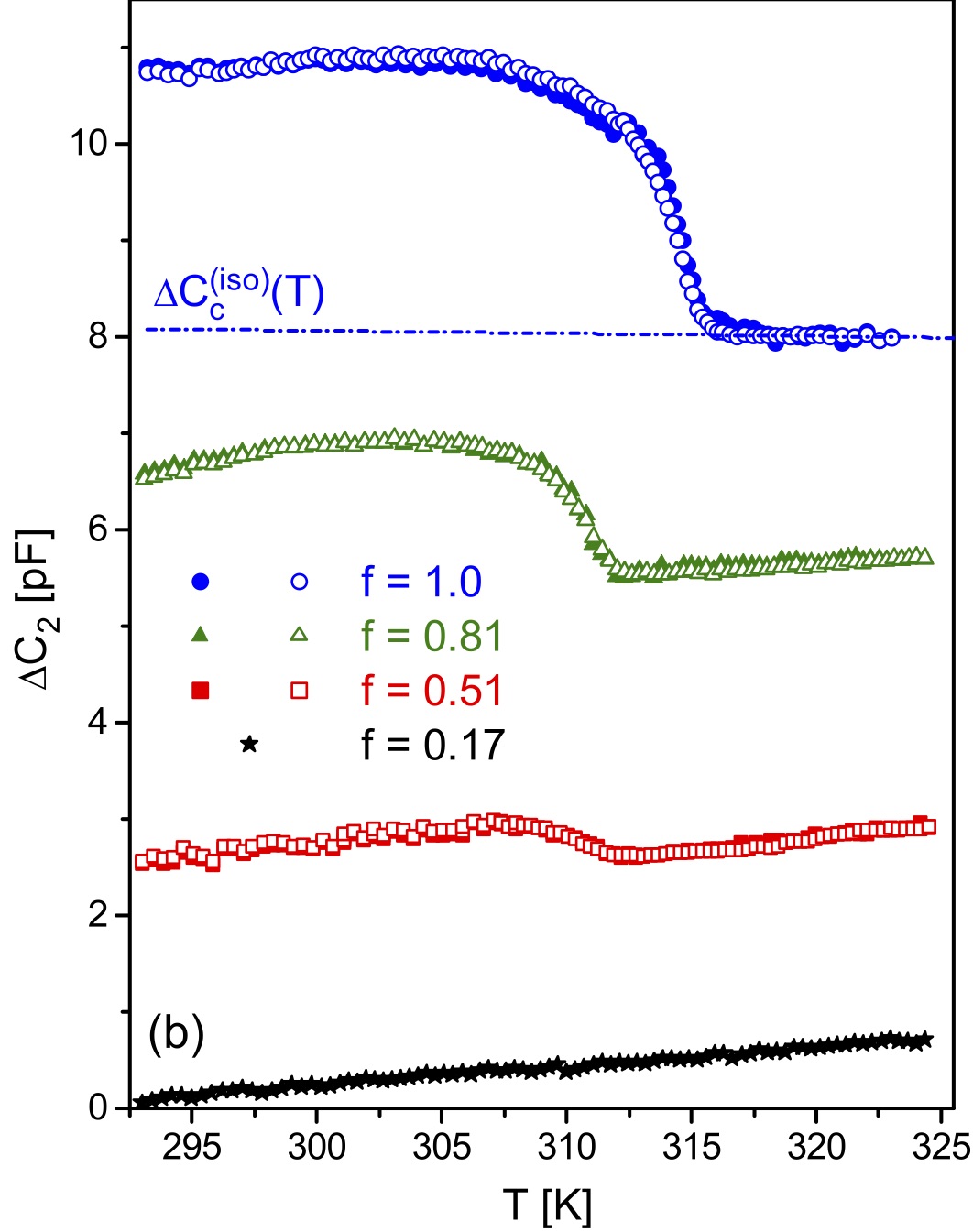}

  \caption{Temperature-dependent capacitance relaxation strength of (a) the slow relaxation process I, $\Delta C_1$ and (b) the fast relaxation process II $\Delta C_2$ for four selected fraction fillings $f$ of 7CB condensed in silica nanochannels. The dash-dotted line is the isotropic baselines, $\Delta C_s^{\mathrm{iso}}(T)$ and $\Delta C_c^{\mathrm{iso}}(T)$ ($f=$1.0) obtained by the linear extrapolation of $\Delta C_1(T)$ and $\Delta C_2(T)$ from the paranematic region, as discussed in the text.} \label{fig:DSRelaxationStrength}
\end{center}
\end{figure}

Based on refractometric data \cite{Chirtoc2004}, we estimated that the relative contribution to the changes of the static capacitance of the composite should not exceed 5\%, which is within the error margins of our fitting analysis. Therefore, the assumption of a temperature-independent $C_{\infty}$ is a good approximation. Moreover, the static dielectric constant, $C_{\mathrm{st}}$, as determined from the Cole-Cole plot by its extrapolation to lower frequencies, does not depend on other extracted fit parameters. Hence, $\Delta C_2$ can be alternatively evaluated by using the equation \ref{eq:CStat}. The magnitudes of $\Delta C_2$, representing the extracted fit parameters and calculated via equation \ref{eq:CStat}, are depicted in Fig.~\ref{fig:DSRelaxationStrength}. In our fitting procedure the parameter set has been optimised in order to achieve a minimal difference between the $\Delta C_2$ values extracted by this alternative calculation procedure. 

Solid lines in Fig.~\ref{fig:DSFillDependence} are the best fits obtained by a simultaneous analysis of the measured real and imaginary parts as outlined above. As a Supplementary the dielectric spectroscopy raw data sets along with fits for several selected temperatures and for all four fraction fillings can be found.

\begin{figure}[H]
\begin{center}
\includegraphics[width=0.7 \linewidth]{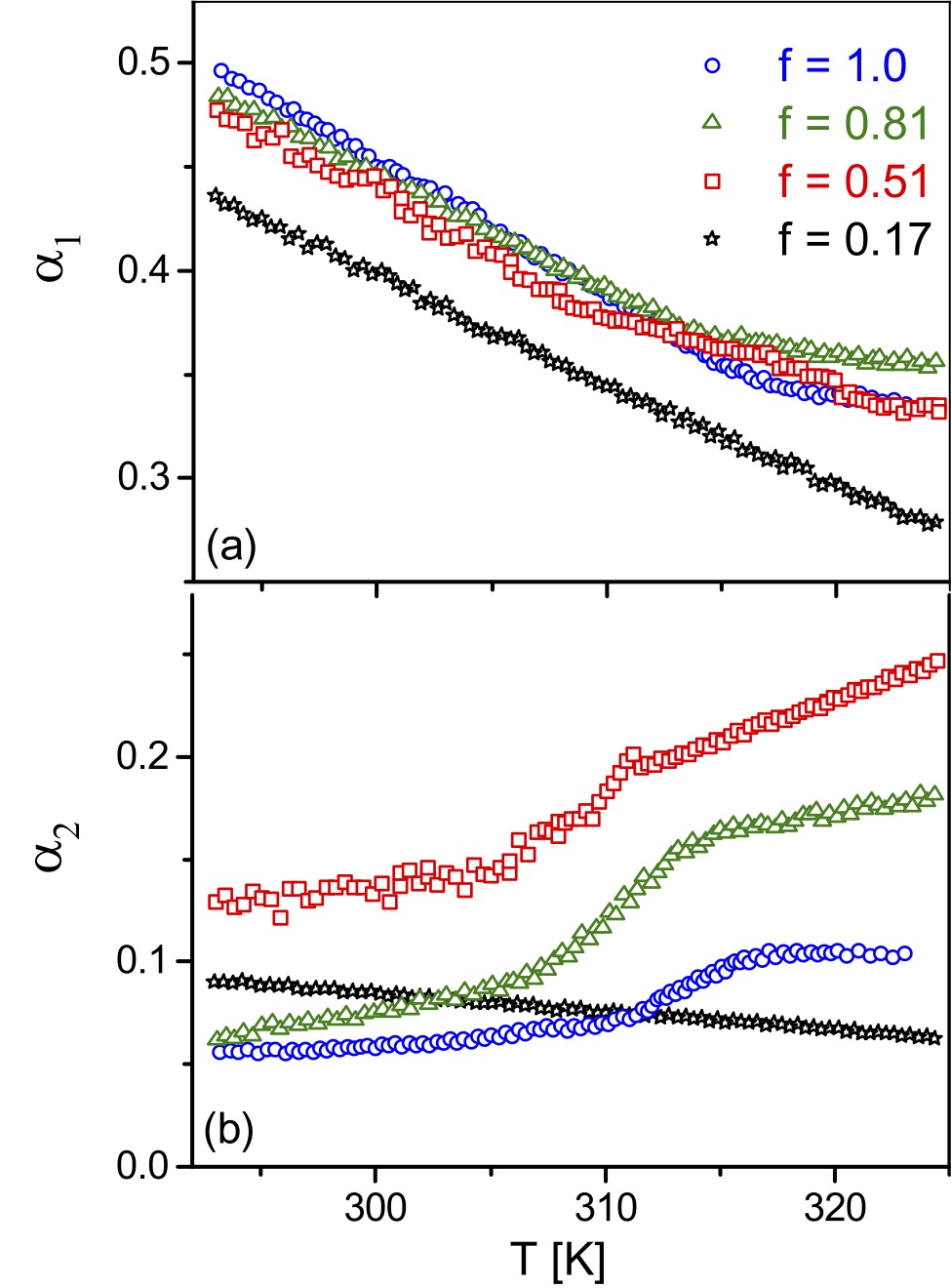}
\includegraphics[width=0.68 \linewidth]{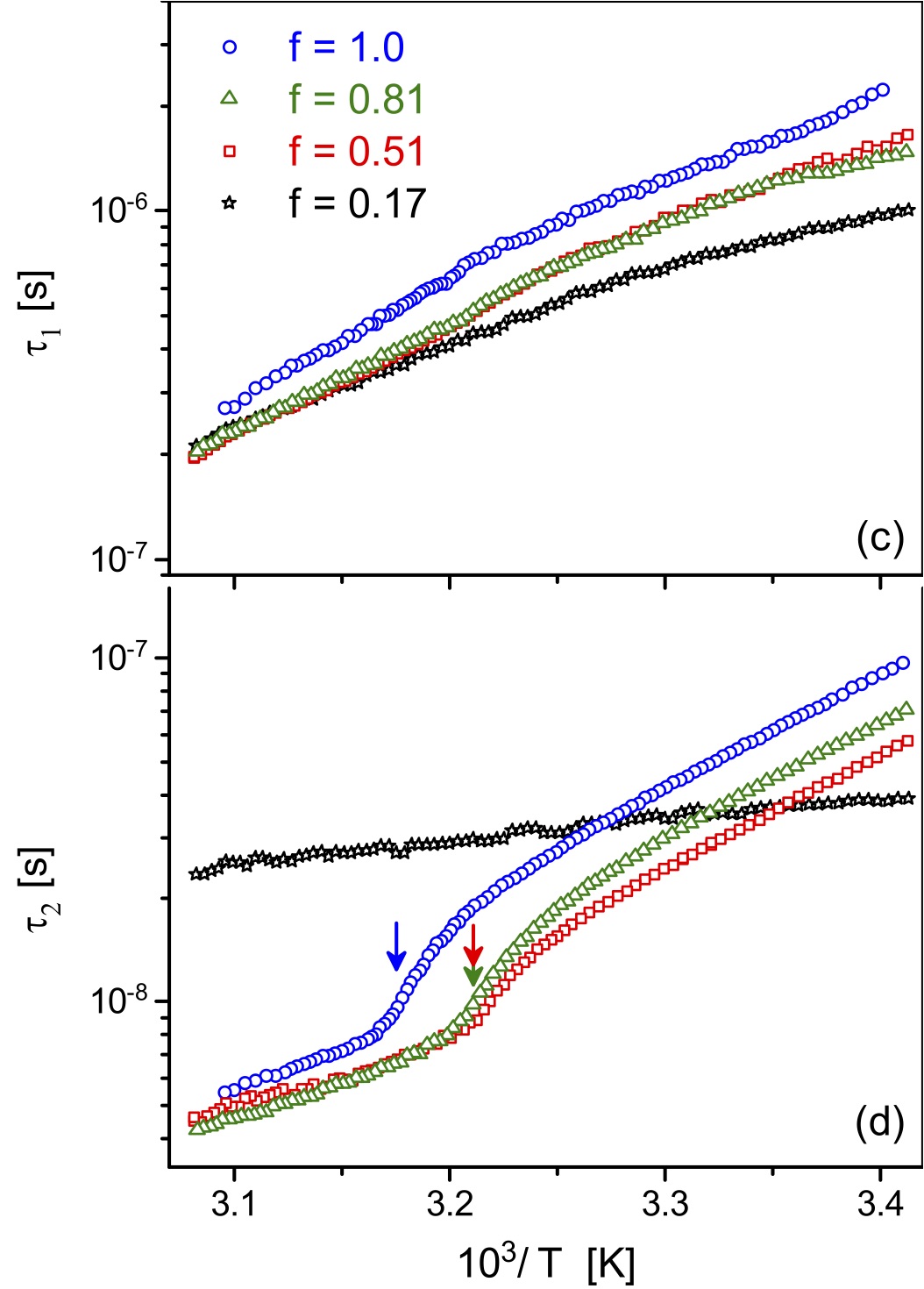}

  

\caption{Temperature-dependent Cole-Cole relaxation distribution parameters of (a) $\alpha_1$, and (b) $\alpha_2$ and the corresponding relaxation times (c) $\tau_1$ (d), $\tau_2$ of the slow and fast relaxation process I and II, respectively, as extracted for four characteristic filling fractions $f$ of silica nanochannels with 7CB. Note that the relaxation times are presented in an Arrhenius plot, \textit{i.e.} on a logarithmic scale versus the inverse temperature.} \label{fig:DSRelaxationTauAlpha}
\end{center}
\end{figure}

The relaxation time of the fast process practically coincides with that of bulk 7CB \cite{Hourri2001}. It is typical of the rotational dynamics around the short molecular axis ($\delta$-relaxation), the end-over-end tumbling\cite{Haws1989,Sinha1998, Diez2006, Frunza2008}.  The $T$ dependences of the extracted fit parameters are depicted in Figs. \ref{fig:DSRelaxationStrength} and \ref{fig:DSRelaxationTauAlpha}. 

In the monolayer regime ($f=$0.17) the dielectric relaxation is dominated by the slow interface process I. However, also a weak contribution of the fast relaxation process II develops as a function of increasing $T$ - see Fig.~\ref{fig:DSRelaxationStrength}b. This behaviour could originate in nonequivalent positions of the nematogen molecules forming the interface layer as a function of $T$. Whereas the filling fraction, $f=0.17$ at $T=293 K$, corresponds approximately to one monolayer, some molecules are expected to be pushed into the next molecular layer at higher $T$s, as sketched in Fig.~\ref{fig:structure}a, because of the increase in thermal activity and intermolecular spacing. As molecular dynamics studies suggest \cite{Ji2009}, such molecules are less spatially constrained and thus exhibit considerably faster relaxations than those being in direct contact with the pore wall (guest-host interaction). Note however that one has to distinguish this fast relaxation process in the film-condensed regime from the one observed in the capillary-condensed regime or in the entirely filled state. Whereas in the first case, the increased mobility is caused by the molecules located at the liquid-vapor interface (Fig.~\ref{fig:structure}a, type 2 molecules), in the second case the fast relaxations are attributable to the molecules in the core region of the pore filling with negligible pore wall/nematogen interaction (Fig.~\ref{fig:structure}, type 3 molecules). This partitioning is corroborated by the smooth $T$ variations of all Cole-Cole parameters of the relaxation process I and II in the film-condensed regime, whereas for higher fractional fillings the relaxation characteristics of the process II are considerably influenced by the P-N transition inside the core region, see Figs.~\ref{fig:DSRelaxationStrength}b and \ref{fig:DSRelaxationTauAlpha}.

In the capillary condensation regime ($f=$0.51 and 0.81) and for the entire filling ($f=$1.0) the Cole-Cole parameters, $\alpha_1$ and $\alpha_2$ (see Fig.~\ref{fig:DSRelaxationTauAlpha}a) exhibit opposite tendencies in their $T$ variation: The interface relaxation I is characterized by a rather broad distribution at room temperature  ($\alpha_1 \sim$ 0.48-0.50). However, it becomes narrower at high $T$s and approaches in the paranematic phase a value 0.36$\pm$0.01. Interestingly, for the fractional fillings, $f\ge 0.51$, the corresponding $\alpha_1(T)$-dependences are all shifted upwards in comparison to the monolayer filling ($f=0.17$). This indicates that by adding molecular layers one modifies also the dynamics of the layer in direct contact with the silica wall. This observation is intuitively understandable and also in good agreement with observations on the reorientational dynamics of methanol molecules upon silica nanopore fillings \cite{Kityk2014}. By contrast, the core relaxation II for the entirely filled sample is characterized by a narrow relaxation rate distribution at room temperature ($\alpha_2 \sim$ 0.06), which broadens in the paranematic state, up to $\alpha_2 \sim$ 0.11. Moreover, as $f$ decreases the $\alpha_2(T)$-curves are systematically shifted upwards. One may speculate that this broadening of the relaxation time distribution can be attributed to molecules located in the menisci regions of the liquid-vapor interfaces (see Fig.~\ref{fig:structure}c and d, - type 3 molecules). Their relaxation properties should differ from the molecules located in the core region, since they encounter more heterogeneous surroundings and thus interaction potentials leading to broader distributions of relaxation times. Also the relative number of these molecules increases with decreasing $f$ \cite{Naumov2008, Page1993, Huber2013}, in agreement with the increased broadening observed for decreasing $f$.


\begin{figure}[H]
\begin{center}
\includegraphics[width=1 \linewidth]{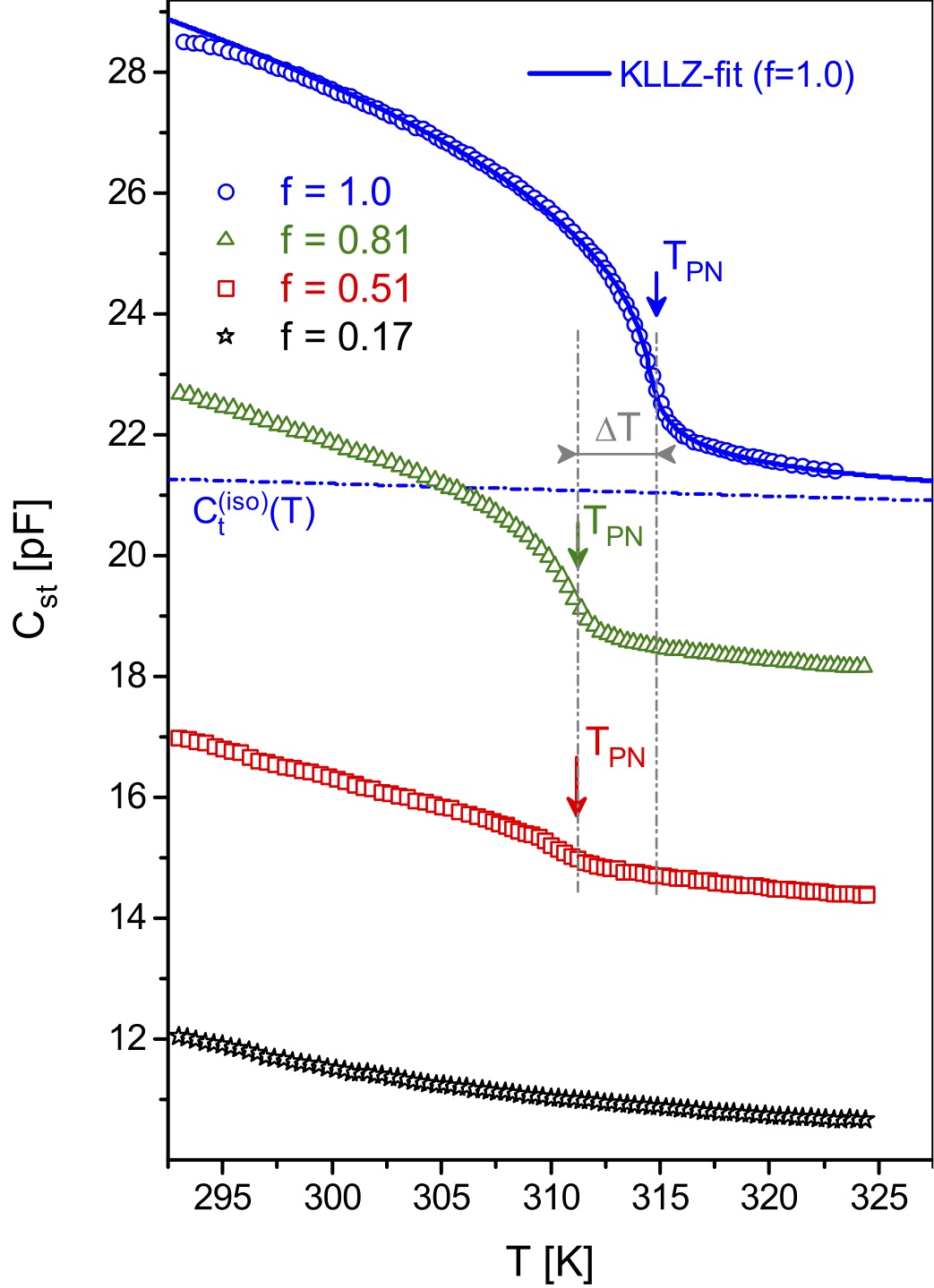}
  \caption{Static capacitance, $C_{\mathrm{st}}$ for four characteristic filling fractions $f$ of silica nanochannels with 7CB. The solid line is the best fit of the $C_{\mathrm{st}}(T)$-dependence ($f=$1.0) based on the KKLZ model. The dash-dotted line is the corresponding baseline, $C_t^{\mathrm{iso}}(T)$, obtained by fitting the KKLZ model.} \label{fig:DSStaticC}
\end{center}
\end{figure}

In order to compare the confined molecular relaxation with the documented molecular dynamics of the bulk LC system, it is instructive to analyse the $T$-dependence of the relaxation times of the interface and core processes, as displayed in Fig.~\ref{fig:DSRelaxationTauAlpha}b. The relaxation times $\tau_1(T)$ and $\tau_2(T)$ rise strongly with increasing orientational ordering. The interface relaxation, however, is much slower than the core relaxation process, which can be traced to the attractive silica potential and the expected larger rotational and translational viscosities, if one has a continuum picture in mind \cite{Sinha1998, Ji2009}. Both in the nematic and in the paranematic phase the fast relaxation in the core region exhibit Arrhenius-like behaviour, $\tau_i=\tau_{oi}\exp(E_{ai}/k_bT)$ ($k_b$ is the Boltzmann constant), typical of a thermally activated dipolar reorientation process across an energy barrier $E_{ai}$.  For the composites with $f=$0.51, 0.81 and 1.0 Arrhenius plots far below $T_{PN}$ yields the activation energies, $E_{o2}$,  64.6, 64.9 and 66.2~kJ/mol, respectively, for the fast, core $\delta$-relaxation process. These values are close to the value of 67.5~kJ/mol reported for bulk 7CB in Ref \cite{Hourri2001}. As expected for the less ordered and thus less constrained state, the activation energies are smaller in the paranematic phase, namely 39.5, 41.0 and 43.0~kJ/mol, respectively. Note, however, that in the vicinity of $T_{PN}$ non-Arrhenius, step-like changes occur (see arrows in the Figure). They can be attributed to the specific relaxation time behaviour, $\tau_{o2}$, caused by the collective orientational molecular ordering at the P-N transition. Overall, the behaviour of $\tau_2(T^{-1})$ is analogous to the characteristic changes of the static capacitance in the vicinity of $T_{PN}$, see Fig.~\ref{fig:DSStaticC}. 

The interface relaxation, $\tau_1$ exhibits a smooth variation accompanied by a gradual slope change, only. There are no sharp features in the vicinity of $T_{PN}$. Obviously, the thermotropic ordering in the interface layer next to the pore wall differs considerably from the one in the core region. This conclusion is also corroborated by the behaviour of the capacitance relaxation strengths, $\Delta C_1(T)$ and $\Delta C_2(T)$, as will be discussed in more detail below. Comparing the slopes of Arrhenius plots $\tau_1(T^{-1})$ in the composites with different fraction fillings, we again see an evident effect of subsequent molecular layers on the mobility characteristic of the interface layer. The Arrhenius plot far below $T_{PN}$ yields the activation energy, $E_{o1}$, of 29.8 kJ/mol for the monolayer which rises with $f$ reaching a magnitude of 45.4 kJ/mol for the composite with $f=1$.

\subsection{Thermotropic nematic order: Landau-de Gennes Analysis and Shell/Core Partitioning}
Both relaxation processes contribute to the static permittivity with considerably different relaxation strengths. As will be outlined in the following this allows us to decompose an effective (average) order parameter behaviour in elementary contributions attributable to molecular rearrangements in the core and in the shell regions, respectively. The orientational order parameter $Q$ of a nematic liquid crystal quantifies the collective preferred molecular orientation in the nematic state. It is defined as $Q=\frac{1}{2}\langle3\cos^2\theta-1\rangle$, where $\theta$ is the angle between the long axis of a 7 CB molecule and the director, the mean direction. The brackets mean an averaging over the molecules under consideration. In general, $Q$ has a local character in a nematic liquid confined in a porous medium, since the orientational ordering near the pore walls and in the core region may differ \cite{Ji2009}. Encouraged by the observation of the two distinct mobility populations, we suggest to approximate the total order parameter, $\bar{Q}$, which characterizes the molecular ordering averaged over the entire pore volume, by a sum of two contributions, an interface (or shell), $\bar{Q_s}$, and a core, $\bar{Q_c}$, contribution:

\begin{equation}
\bar{Q}=\bar{Q_s}+\bar{Q_c}=w_s Q_s + w_c Q_c. \label{eq3}
\end{equation}

Here the elementary contributions, $\bar{Q_s}$ and $\bar{Q_c}$, are defined by local order parameters $Q_s$ and $Q_c$ and weighted according to their volume fractions, $w_s$ and $w_c$ with $w_s + w_c=1$. Note that $Q_s$ and $Q_c$ are assumed to be constant over the distinct pore regions.

The relation between the static dielectric permittivity and the orientational order parameter $Q$ is given by the Mayer and Meier equation \cite{Maier1961}. Hence, for a nematic molecule having the dipole moment nearly parallel to the molecular long axis and with the director aligned parallel to the probing field, the capacitance relaxation strength, $\Delta C$ can then be expressed in a quite simple way:
\begin{equation}
\Delta C=\Delta C^{\mathrm{iso}}+p Q, \label{eq5}
\end{equation}
where the first term and the factor of proportionality, $p$, depend on the molecular dipole moment and its orientation with respect to the long principal axis, the molecular number density, the internal molecular fields and the temperature. Both $\Delta C^{\mathrm{iso}}$ and $p$ are in principle $T$ dependent, but in most cases the corresponding changes are weak, particularly if the relative changes in the absolute temperature are small, as is the case here. For the bulk nematic  LC $\Delta C^{\mathrm{iso}}(T)$ represents the bare $T$ dependence of the capacitance relaxation strength in the isotropic phase, whereas its extrapolation to the nematic phase gives an \emph{isotropic baseline} relative to which an excess contribution due to orientational ordering ($p$-term) occurs. The Maier-Meier equation can be applied in principle to each contribution separately. Thus, the static capacitance is given as
\begin{equation}
C_{\mathrm{st}}=C_t^{\mathrm{iso}}+p(\bar{Q_s}+\bar{Q_c})= C_t^{\mathrm{iso}}+p\bar{Q},\label{eq:CDecomp}
\end{equation}
where $C_t^{\mathrm{iso}}(T)= C_{\infty}+\Delta C_s^{\mathrm{iso}}(T)+\Delta C_c^{\mathrm{iso}}(T)$ is the isotropic baseline of static capacitance, $\Delta C_s^{\mathrm{iso}}(T)$ and $\Delta C_c^{\mathrm{iso}}(T)$ are the isotropic baselines of shell and core relaxation strengths, respectively. 

Note that in the paranematic state ($T>T_{PN}$) $C_{\mathrm{st}}\ne C_t^{\mathrm{iso}}$, because of a residual nematic ordering in the interface region. In the presented approach the excess capacitance, $C_{\mathrm{st}}(T)- C_t^{\mathrm{iso}}(T)$ is proportional to the effective order parameter, $\bar{Q}$, and will be subjected to a further analysis.

\begin{figure}[H]
\begin{center}
\includegraphics[width=1 \linewidth]{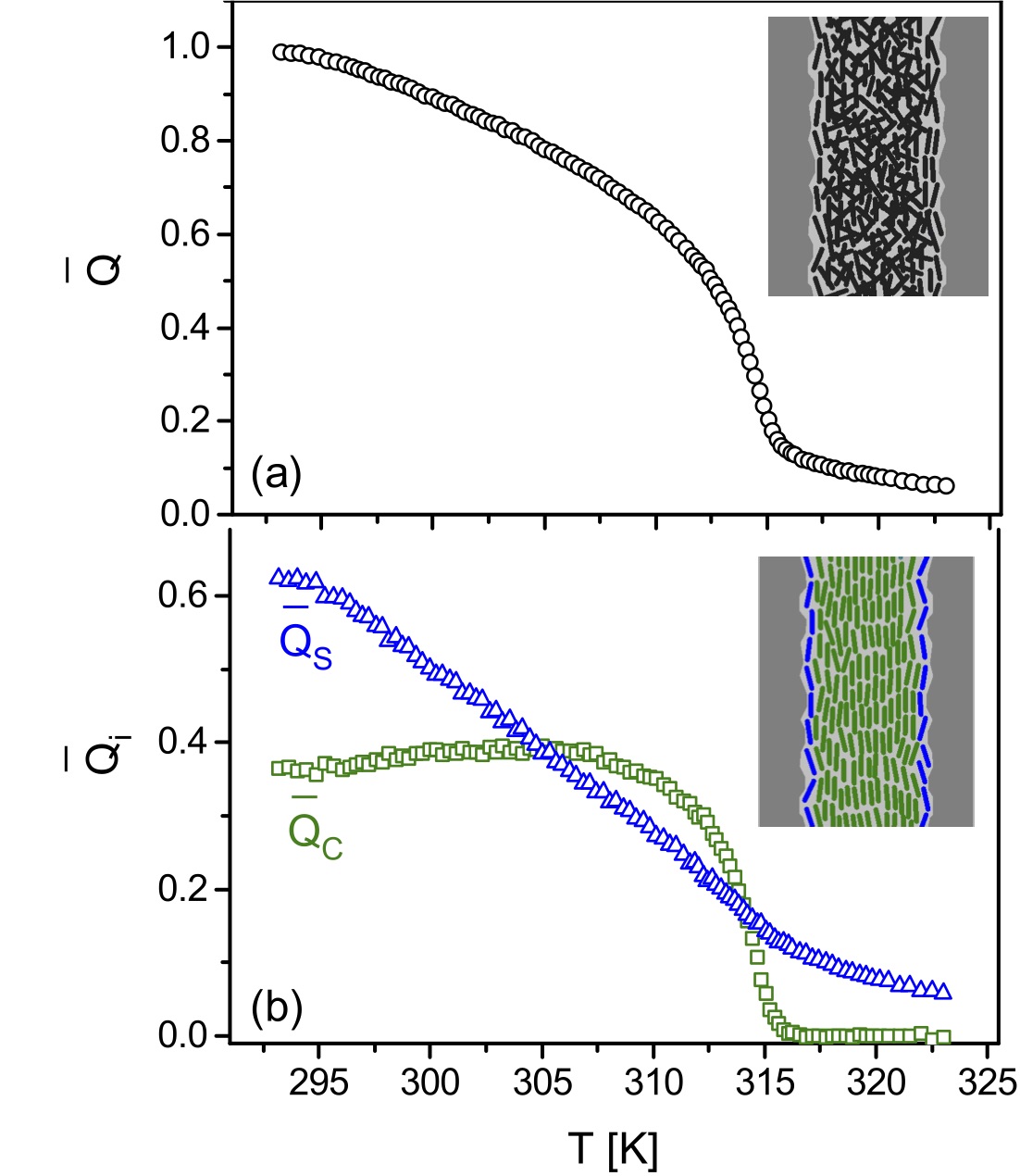}
 \caption{The effective (averaged) orientational order parameter, $\bar{Q}$ vs $T$ (a) and its deconvolution on the elementary local order parameters describing molecular orderings in the shell ($\bar{Q_s}(T)$) and core ($\bar{Q_c}(T)$) regions (b) - see also the colored illustration of the different molecular populations. For convenience, all order parameters are normalized to the value $\bar{Q}$ at 293 K. The relation $\bar{Q}=\bar{Q_s}+\bar{Q_c}$ holds for each $T$.}\label{fig:OrderParameterDeconvolution}
\end{center}
\end{figure}

The difference, $C_{\mathrm{st}}(T)- C_t^{\mathrm{iso}}(T)\propto \bar{Q}$ normalized to a value at $T=293$ K, $\bar{Q}(T)$-dependence is depicted in Fig.~\ref{fig:OrderParameterDeconvolution}a. The effective (averaged) order parameter $\bar{Q}$ can be deconvoluted into elementary contributions characterizing the molecular orderings in the interface shell ($\bar{Q_s}$) and core ($\bar{Q_c}$) regions:
\begin{eqnarray}
\bar{Q_s}(T)\propto \Delta C_1(T) -  \Delta C_s^{\mathrm{iso}}(T); \\ \nonumber
 \bar{Q_c}(T)\propto \Delta C_2(T) -  \Delta C_c^{\mathrm{iso}}(T). \label{eq:QDecomp}
\end{eqnarray}
where $\Delta C_1(T)$ and $\Delta C_2(T)$ are the capacitance relaxation strengths of the slow and fast processes, respectively.  $\Delta C_s^{\mathrm{iso}}(T)$ and $\Delta C_c^{\mathrm{iso}}(T)$ are the isotropic baselines. $\Delta C_c^{\mathrm{iso}}(T)$ can be easily determined, if one takes into account that $\Delta C_2(T)$ saturates to nearly $T$ independent values immediately above the nematic-to-paranematic transition point, $T_{PN}$. We believe that a linear extrapolation of this dependence below $T_{PN}$, as it is depicted in Fig.~\ref{fig:DSStaticC}a (see dash-dot line), properly describes the corresponding baseline in the confined nematic phase. $\Delta C_1(T)$, on the other hand, evidently exhibits a nonlinear asymptotic behaviour which extents far above $T_{PN}$. Nevertheless, the baseline can be determined, if one considers Eq. \ref{eq:CDecomp}: $\Delta C_s^{\mathrm{iso}}(T)= C_t^{\mathrm{iso}}(T)-C_{\infty}-\Delta C_c^{\mathrm{iso}}(T)$, see the dash-dot line in Fig.~\ref{fig:DSRelaxationStrength}(a)





In Fig.~\ref{fig:DSStaticC} we present the $T$ dependences of the static capacitance for the composites for all $f$s investigated. In the monolayer regime ($f=0.17$), there is only a gradual increase of $C_{\mathrm{st}}$ upon cooling, indicating a weak increase in orientational ordering. At higher $f$s pronounced changes with a characteristic kink at $T_{PN}$ in the  $C_{\mathrm{st}}(T)$ dependences are observed. Their \emph{continuous} evolution (with a precursor behaviour at high $T$s) contrasts with the discontinuous behaviour observed at the isotropic-to-nematic transition of bulk 7CB \cite{Kityk2008}.

More quantitative insights in this phenomenology can be achieved by a comparison with the KKLZ-model \cite{Kutnjak2003,Kutnjak2004} for the isotropic-to-nematic transition in cylindrical pore geometries  briefly discussed in the introduction. According to the KKLZ approach the orientational ordering in confinement is characterized by the reduced order parameter $q=\bar{Q}/Q_0$, where $Q_0=Q(T_{\rm IN})$ is the bulk value of the order parameter taken at the isotropic-to-nematic transition temperature, $T_{\rm IN}$. The dimensionless Landau-de Gennes free energy of the nematic LC is then represented as
\begin{equation}
f = t q^2 - 2 q^3 + q^4 - q \sigma + \kappa q^2, \label{eq:KKLZ}
\end{equation}
where $t=(T-T^*)/(T_{IN}-T^*)$ is the dimensionless temperature \cite{Kutnjak2003}. A crucial parameter of the model is the difference $T_{\rm IN}-T^*=\Delta T^*$, which sets the ''effective" temperature $T^*$ and thus calibrates the $T$-scale of the KKLZ-model. Based on the bulk birefringence behaviour in the isotropic-to-nematic transition \cite{Calus2014}, $\Delta T^*$ was determined to 4.9~K. A bilinear coupling between the order parameter and the nematic ordering field, \textit{i.e.} the $q \sigma$-term, is a key feature of the KKLZ theory.  It results in an upward shift of the effective transition temperature, $T_{PN}$, and a residual nematic ordering at $T>T_{PN}$, typical of the paranematic state, as a function of increasing strength of the interface ordering field, $\sigma$. A $\sigma$-value of $\sigma_c=0.5$ corresponds to a critical point in the $\sigma-T$ phase diagram, which separates lines of continuous and discontinuous phase transition evolution \cite{Kutnjak2003}. The $\kappa$-term accounts for quenched disorder effects attributable to static variations of $\sigma$, here most prominently attributable to pore wall inhomogeneities \cite{Kutnjak2003, Kityk2008}. It rises the free energy of the low temperature phase and thus provides, in contrast to a finite value of $\sigma$, a downward shift of the effective transition temperature.

The KKLZ model is applicable to entire filled channels only, since the orientational distortions occurring at the menisci interfaces for the partially filled state are not considered. In addition, our experimental results are affected by depolarization effects at the menisci interfaces. Therefore, we restricted our analysis to the entirely filled samples ($f=1$). Minimization of Eq. \ref{eq:KKLZ} with respect to $q$ gave the equilibrium value of the scaled order parameter, $q_e$, which can be converted to the $T$-dependence of the static capacitance and compared with experiment. By an iterative least-square numerical fitting procedure a very good agreement between experiment and theory was obtained for $\sigma$=0.76 and $\kappa=$0.96, see solid curve in Fig. \ref{fig:DSStaticC}. It quantitatively accounts for both the paranematic behaviour above, and the continuous transition behaviour at  $T_{PN}$. According to the KKLZ model $\sigma$ is proportional to the inverse pore radius, $R^{-1}$. This allows one to calculate a theoretical, critical channel radius, $R_c=R \sigma/\sigma_c$ \cite{Kutnjak2003} which would separate a continuous (for $R<R_c$) from a discontinuous P-N transition (for $R>R_c$). It corresponds to 10~nm, a value in agreement with the one experimentally determined from channel-radius dependent birefringence measurements on 7CB \cite{Calus2014}.

Another interesting feature in Fig. \ref{fig:DSStaticC} is a $T$-shift of the transition temperate, $T_{PN}$ of $\Delta T \sim$3.5 K, between the capillary-condensed fillings ($f=$0.51, 0.81) and the entirely filled channel  ($f=$1). One may speculate that competing anchoring forces and elastic deformations at the liquid-vapor menisci interfaces in the partially filled state (as sketched in Fig.~\ref{fig:structure}c) could be responsible for this stronger preference of the disordered state in the partial fillings. There is, however, another, quantitatively verifiable and arguably more simple explanation, which is related to the hydrostatics of the confined liquid: In the partially filled, capillary-condensed state, the liquid experiences a tensile pressure, dictated by the concave curvature of the menisci terminating the liquid bridges. This negative pressure causes not only subtle deformations of the rigid, nanoporous matrix \cite{Guenther2008, Prass2009, Gor2010}, it also significantly affects density, and thus pressure-dependent phase transformations, most prominently the liquid-solid transition \cite{Morishige2006, Schaefer2008, Morishige2012, Moerz2012}, but also the isotropic-to-nematic transition studied here \cite{Rein1993, Manjuladevi2002}. According to the Young-Laplace formula applied to the menisci in the capillaries (mean curvature radius = - pore radius = -6.6 nm, surface tension of 7 CB=31.7~mN/m \cite{Delabre2009}) the tensile pressure in the liquid bridges amounts to $\sim$-9.6 MPa for $f_{\rm c}<f<1$. This pressure is completely released upon reaching $f=1$ (mean curvature radius = $\infty$). An extrapolation of $T_{\rm IN}$ of bulk 7CB as a function of positive, compression pressure, reported in the literature \cite{Rein1993}, towards this magnitude of negative, tensile pressures yields an expected, 3.5~K Young-Laplace pressure induced downward shift of the transition, in excellent agreement with the $\Delta T$ observed. Therefore, we think that the $T$-shift is rather another vivid manifestation of the high tensile pressures in the capillary-condensed state, than attributable to competing anchoring conditions. This conclusions is further corroborated by the observation that the core relaxation time $\tau_2$ is evidently shorter in the capillary-condensed state than in the completely filled state, see Fig. \ref{fig:DSRelaxationTauAlpha}. This faster molecular reorientation relaxation is also compatible with a negative Young-Laplace pressure. It results in a reduced density in the liquid bridges, compared to the un-stretched liquid, and thus in a reduced rotational viscosity of the confined liquid.

Given the additivity in the Maier-Meier equation for the static dielectric relaxations and the resulting orientational order, we discuss in the following a superposition of the orientational order in a shell and in a core contribution under a consideration of a proper treatment of the $T$-dependent baselines of the static capacitance contributions (see Methods). The resulting $T$-dependent behaviour of the order parameter averaged over the silica nano channels and normalised to its value at $T=$293~K, $\bar{Q}(T)$ is depicted in Fig.~\ref{fig:OrderParameterDeconvolution}a. It can be decomposed into elementary contributions, calculated by the base-line corrected relaxation strengths, which are characteristic of the molecular orderings in the shell ($\bar{Q_s}$) and core ($\bar{Q_c}$) regions, as displayed in Fig.~\ref{fig:OrderParameterDeconvolution}b. The $T$ variations of the different components in the proximity of $T_{PN}$ differ considerably, indicating a distinct thermotropic orientational behaviour: Whereas the core ordering exhibits an abrupt and almost jump-like increase in nematic ordering, reminiscent of the bulk behaviour, the interface ordering shows continuous changes with a "rounded" kink near $T_{PN}$ and an asymptotic decrease above this temperature. 

It is understood that this linear decomposition is a simplification, for it assumes an entirely decoupled thermotropic behaviour between shell and core. There is, however, evidence that this is not strictly true: (i) We saw changes in the relaxation dynamics of the slow interface process as a function of $f$. (ii) The core order parameter exhibits a small, but systematic decrease with $T$ below 305~K. This peculiarity could be an artefact of the decomposition, but it also could be real. For example, the increase in $Q_i$ (and the related stiffening) in the interface layer may lead to an increased impact of the pore wall irregularities on the thermotropic order in the core, and thus to a decrease of $Q_s$. (iii) As one can learn from the $\Delta C_1(T)$-dependences, typical of the order parameter $Q_s$, the slope of the dependences above $T_{PN}$ are nearly the same for all $f$s, see Fig.~\ref{fig:DSRelaxationStrength}a. However, below $T_{PN}$ $\Delta C_1(T)$ remains unchanged for the composite with a monolayer (or shell) filling ($f=0.17$), only, and obviously exhibits an increasing slope as a function of increasing $f$. This indicates, that the paranematic ordering in the host-guest interface evolves independently solely for the monolayer filling. By contrast, upon additional channel filling the thermotropic  ordering in the interface layer is enhanced by thermotropic ordering processes of the core, presumably via intermolecular interactions. Thus, we believe that the distinct behaviour of the relaxation strength and relaxation dynamics (broadening and relaxation times) of the two processes well justify a qualitative partitioning in a core and a shell behaviour, there is, however, still a final interaction between the two molecular populations. It is interesting to note that such a coupling is also found in Molecular Dynamics simulations by Ji \textit{et al.} on rod-like LCs in tubular channels \cite{Ji2009}, even though their radially resolved order parameter and relaxation time profiles suggest an even stronger mutual influence of the shell and the core behaviour as observed here.

\section{Conclusions}

We reported dielectric studies on the nematic crystal 7CB confined in parallel-aligned nanochannels in monolithic silica membranes. The measurements have been performed on composites characterized by different fractional fillings that cover film condensation at the channel wall, capillary condensation and the entire filling. Whereas for the composite with a LC film a slow interface relaxation dominates the dielectric properties, in all other cases the dielectric spectra can be well described by two thermally activated Cole-Cole processes with considerably different relaxation rates, \textit{i.e}. (i) a slow relaxation at the interface region in the LC layer next to pore walls and (ii) a fast relaxation in the core of the nano channel filling. 

{
Our findings compare well with experimental results on the dynamics of "conventional" liquid condensates \cite{Farrer2003, Valiullin2004, Morineau2004, Knorr2008, Schranz2007, Hofmann2012, Kityk2014, Huber2015} confined in nanoporous media. In particular, the observation of distinct core and shell molecular mobilities is also in good agreement with experimental studies on pore-confined liquid crystalline systems presented in the past  \cite{Crawford1996, Cramer1997, Hourri2001, Frunza2001, Leys2005, Sinha2005, Leys2008, Frunza2008, Bras2008, Aliev2005, Aliev2010, Jasiurkowska2012}, most prominently with the work by Aliev \etalp discussed in the introduction. Going beyond this work, we demonstrate here that the core ordering is reminiscent of the abrupt, discontinuous bulk order behaviour, whereas the surface ordering exhibits a much more continuous evolution with a gradual change in slope at the P-N transition point. It is asymptotically decreasing above this temperature and the superposition of both behaviours results in an effective isotropic-to-paranematic transition which can be well described by a phenomenological Landau-de Gennes Ansatz (KKLZ-model).} 

{These experimental results corroborate the radial partitioning and gradual thermotropic orientational ordering behaviour found in computer simulations for rod-like LCs in cylindrical nanochannels. \cite{Ji2009, Guegan2007, Lefort2008, Karjalainen2015} Albeit the simulation studies indicate a rather smooth transition in the molecular mobility  \cite{Ji2009, Guegan2007, Lefort2008} between the interface region and the core, whereas according to our experiments this transition is quite sharp, resulting in a very distinct monolayer and core behaviour. Moreover, our filling-fraction dependent experiments provide information on the thermotropic orientational order as a function of mesoscopic arrangement of the nematic liquid in the channels (capillary- and film-condensed states). These different fillings states have not been explored in simulation studies so far.} 

For the future, it would be interesting to what extent computer simulations can shed additional light on the impact of the tensile pressure in the capillary-condensed state in comparison to the influence of possible heterogeneous anchoring conditions in this state of the confined liquid. In fact, the advent of high brilliant Synchrotron-based X-ray sources may allow one to perform already in the near future single-channel experiments in order to explore spatially resolved anchoring profiles and the intimately related \textit{translational} liquid crystalline order. Such structural information would be of high importance in order to complement the dynamic and static information regarding the \textit{orientational} order of highly spatially confined nematic liquids unravelled here by dielectric spectroscopy experiments on arrays of nanochannels.

Supporting Information Available:
The frequency dispersions of the complex capacitance for seven selected temperatures and four fraction fillings of 7CB embedded into silica nanochannels ($R=6.6$ nm) is available free of charge via the Internet as RSC's Electronic Supplementary Information (ESI).

\section{Acknowledgement}

This work has been supported by the Polish National Science Centre (NCN) under the Project "Molecular Structure and Dynamics of Liquid Crystals Based Nanocomposites" (Decision No. DEC-2012/05/B/ST3/02782). The German research foundation (DFG) funded the research by the research Grant No. Hu850/3 and within the collaborative research initiative "Tailor-made Multi-Scale Materials Systems" (SFB 986, project C1), Hamburg.

\bibliographystyle{unsrt}


\end{document}